\documentclass[aps,prd,showpacs,nofootinbib,floatfix,twocolumn]{revtex4}
\usepackage{amsmath,amssymb,bm,epsfig}
\usepackage{slashed}

\newcommand\pd{\partial}
\renewcommand\Re{{\rm Re\,}}
\renewcommand\Im{{\rm Im\,}}

\newcommand\<{{\langle}}
\renewcommand\>{{\rangle}}
\renewcommand\O{{O}}
\newcommand\U{{\cal U}}

\def\Xint#1{\mathchoice
   {\XXint\displaystyle\textstyle{#1}}%
   {\XXint\textstyle\scriptstyle{#1}}%
   {\XXint\scriptstyle\scriptscriptstyle{#1}}%
   {\XXint\scriptscriptstyle\scriptscriptstyle{#1}}%
   \!\int}
\def\XXint#1#2#3{{\setbox0=\hbox{$#1{#2#3}{\int}$}
     \vcenter{\hbox{$#2#3$}}\kern-.5\wd0}}
\begin{document}
\title{Deconstruction of Unparticles}
\author{M.~A.~Stephanov}
\affiliation{Department of Physics, University of Illinois, 
Chicago, Illinois 60607-7059,USA}

\begin{abstract}
We discuss properties of hypothetical scale invariant (unparticle)
matter by viewing it as a tower of massive particles.
We show how peculiar properties of unparticles emerge in the
limit when the mass spacing parameter $\Delta$ vanishes. 
We explain why unparticle
cannot decay in this limit and how, for finite $\Delta$, the decays
manifest themselves in a relation between the reconstructed invariant mass
and vertex displacement. We describe a model field theory in AdS$_5$
which explicitly implements the deconstruction procedure by truncating
the extra dimension to size of order $1/\Delta$.
\end{abstract}
\date{May 2007}
\pacs{}

\maketitle


\section{Introduction}

In a recent paper \cite{Georgi:2007ek} Georgi suggests
to consider 
a hypothetical scale invariant, or
conformal, matter very weakly coupled to the Standard Model matter
as a possible component of physics above TeV scale which might begin
to show up at the LHC.
Vast amount of knowledge exists about the properties of conformal
field theories, and its applications to condensed matter problems,
such as critical phenomena, abound. However, there is little
understanding of how such a conformal sector, if it exists, would
manifest itself in particle physics experiments. Georgi terms such
matter ``unparticle'', for its properties and signatures are 
qualitatively different
from those of particles. Several curious properties of
unparticles have been exposed in
Ref.~\cite{Georgi:2007ek,Georgi:2007si}, including the unusual scaling
of the apparent phase space volume, the unusual missing energy
spectra,
unusual interference patterns, etc.  Several recent papers have been
addressing  novel signatures of
unparticles~\cite{Cheung:2007ue,Luo:2007bq,Chen:2007vv,Ding:2007bm,Liao:2007bx,Aliev:2007qw,Li:2007by}.

The purpose of this paper is to clarify the notion of the unparticle
using the language familiar to a particle physicist. To that end we
deconstruct the unparticle and view it
as an infinite tower of particles of different masses.%
\footnote{We rely on a more generic meaning of the term
  ``deconstruction'', and wish to emphasize the difference from
so far more common specific use of this term to describe deconstruction of
  5-dimensional theories by discretization of the 5-th dimension.}
We show how the peculiar properties of unparticles exposed in
Refs.~\cite{Georgi:2007ek,Georgi:2007si} can be interpreted and
understood in that language. 

We shall think of the unparticle as a limiting case in which the
spacing  $\Delta^2$ of the (squared) masses in the tower of particles 
goes to zero. Using
this limiting procedure we explain an apparent paradox mentioned in
Ref.~\cite{Georgi:2007si}, that unparticle need not decay despite the
presence of finite imaginary part in its two-point correlator.
As a corollary we find that if $\Delta$ is small but finite, the
unparticle can decay, and we describe peculiar signatures of such
decays.

In the beginning we shall view
deconstruction as a purely mathematical device translating the
properties of unparticle into particle physics language. In
Section~\ref{sec:model-deconstr-unpar} we shall take more constructive
approach and describe a model of a field theory in which the
unparticle indeed arises as a limiting procedure. Not surprisingly,
the model requires an additional space dimension, and the tower
of deconstructing particles appears naturally as a Kaluza-Klein tower,
once the extra dimension is compactified/truncated. In order to
produce an unparticle with non-integer scaling dimension we use
AdS$_5$ geometry and describe unparticle using a massive scalar field.

\section{Setup and notations}
\label{sec:setup-notations}

Following Ref.\cite{Georgi:2007ek}, let us imagine that there exists a
scale invariant sector of our world described, e.g., by some strongly
selfcoupled conformal theory. Scale invariance means that there
are no (massive) particles in this sector. Since such unparticle
sector is not seen in present experiments, we must further assume that
the coupling of this conformal sector to the Standard Model
particles/fields is very weak -- somewhat in the spirit of the
``hidden-valley'' models~\cite{Strassler:2006im}. This interaction can
be described in an effective field theory language as a
(non-renormalizable) coupling between some Standard Model operator and
the unparticle operator $\O$ of scaling dimension $d_\U$. Since the
unparticle sector is selfinteracting, the dimension of
$\O$ can be nontrival (noninteger).

The correlation function of the operator is given by
\begin{equation}
  \begin{split}
  \label{eq:OO}
  &\int d^4x e^{iPx}\<0|T\O(x)\O^\dag(0)|0\> \\
&= \int \frac{dM^2}{2\pi} \rho_\O(M^2) \frac{i}{P^2-M^2+i\varepsilon}
\end{split}
\end{equation}
By scale invariance the spectral function of the operator~$\O$ must
be a power of $M^2$:
\begin{equation}
  \label{eq:rhoO-scaling}
  \rho_\O(M^2) = A_{d_\U} (M^2)^{d_\U-2},
\end{equation}
where $A_{d_\U}$ is a normalization constant chosen by convention in 
Ref.\cite{Georgi:2007ek}. Its precise form/value is not consequential
for the discussion below. On the other hand
\begin{equation}
  \label{eq:rhoO}
  \rho_\O(M^2) = 2\pi \sum_\lambda \delta(M^2-M_\lambda^2) |\<0|\O(0)|\lambda\>|^2.
\end{equation}
The sum in Eq.~(\ref{eq:rhoO}) is over all relativistically normalized
states $|\lambda\>$ 
(at fixed spatial momentum -- i.e., no $d^3p_\lambda$ integration).
The unparticle spectral function means that the spectrum of the
operator $\O$ is continuous, i.e., the sum in Eq.~(\ref{eq:rhoO})
is in fact an integral. Let us imagine that the scale invariance is
broken in the system in a controllable way, so that, instead of a
continuous spectrum of states $\lambda$, there is a discrete tower of states
with the spacing controlled by parameter $\Delta$. To simplify discussion,
we shall pick a particular spectrum
\begin{equation}
  \label{eq:Mn}
  M_n^2=\Delta^2 n.
\end{equation}
It is straightforward to adjust all subsequent discussion to any other
spectrum, e.g., $M_n^2=\Delta^2 n^2$, etc. We shall assume that
$\Delta$ is much smaller than other scales pertinent to the problem.
Let us introduce notation for the matrix element for the $n$th particle:
\begin{equation}
  \label{eq:Fn}
  F_n^2\equiv |\<0|\O(0)|\lambda_n\>|^2
\end{equation}
We can then write:
\begin{equation}
  \label{eq:rho-Fn}
    \rho_\O(M^2) = 2\pi \sum_n \delta(M^2-M_n^2) F_n^2.
\end{equation}
and 
\begin{equation}
  \label{eq:OO-Fn}
    \int d^4x e^{iPx}\<0|T\O(x)\O^\dag(0)|0\> 
= \sum_n \frac{iF_n^2}{P^2-M_n^2+i\varepsilon}
\end{equation}
In the limit $\Delta\to0$ the sum over $n$ in Eq.~(\ref{eq:rho-Fn})
becomes an integral, which must match Eq.~(\ref{eq:rhoO-scaling}).
From this condition we easily determine that $F_n$ must be given by
\begin{equation}
  \label{eq:Fn-Delta}
  F_n^2 = \frac{A_{d_\U}}{2\pi}\Delta^2(M_n^2)^{d_{\U}-2}
\end{equation}
The constants $F_n$ are similar to the decay constants of mesons in
QCD. 
More generally:
\begin{equation}
  \label{eq:Fn-gamma}
  M_n^2=\Delta^2n^{1/\gamma}
\quad\Rightarrow\quad
F_n^2 = \frac{A_{d_\U}}{2\pi\gamma}\Delta^{2\gamma}(M_n^2)^{d_{\U}-2-(\gamma-1)}
\end{equation}

\section{Production}
\label{sec:production}

Consider example from Ref.~\cite{Georgi:2007ek} of production
of such an unparticle.
Imagine the coupling of the unparticle given by
\begin{equation}
  \label{eq:utU}
  i\frac{\lambda}{\Lambda_\U^{d_\U}}\bar
  u\gamma_\mu(1-\gamma_5)t\,\pd^\mu\O+{\rm h.c.}
\end{equation}
where $\lambda$ is a dimensionless coupling and $\Lambda_\U$ is the
Banks-Zaks scale in the unparticle theory. Using
representation~(\ref{eq:OO-Fn}) we can define the deconstructing
particle field
\begin{equation}
  \label{eq:lambda_n-O}
  \lambda_n(x)\equiv \O(x)/F_n.
\end{equation}
 According to (\ref{eq:OO-Fn}), on the mass shell of the $n$'s particle
this field will be canonically normalized. Thus, the interaction
Eq.~(\ref{eq:utU}) becomes after deconstruction
\begin{equation}
  \label{eq:lambda_n}
    i\frac{\lambda}{\Lambda_\U^{d_\U}}\bar
  u\gamma_\mu(1-\gamma_5)t\,\sum_n F_n \pd^\mu\lambda_n+{\rm h.c.}
\end{equation}
Now it is easy to study the production of the unparticle using the
standard notions of the particle physics. The kinematics is that of a
two body decay of a $t$ quark. That is, for each $n$ the energy of
the $u$ quark is fixed to $E_u=(m_t^2-M_n^2)/(2m_t)$. The spectrum of
$E_u$ consists of a peak for each value of $n$, which in the limit
$\Delta\to\infty$ merge into the continuum distribution displayed in
Ref.~\cite{Georgi:2007ek}. The decay rate for each $n$ is
\begin{equation}
  \label{eq:dGamma}
  \Gamma(t\to u+\lambda_n) = \frac{|\lambda|^2}{\Lambda_\U^{2d_\U}}
\frac{m_t E_u^2}{2\pi}\,F_n^2.
\end{equation}
The interval $dE_u$ corresponds to the interval of masses
$dM^2=2m_tdE$ which contains $2m_tdE/\Delta^2$ states $\lambda_n$.
Thus we obtain
\begin{equation}
  \begin{split}
  \label{eq:dGamma-dE}
  &\frac{d\Gamma}{d E_u} = \frac{2m_t}{\Delta^2}\Gamma(t\to
  u+\lambda_n)\\
&= \frac{2m_t}{\Delta^2} \frac{|\lambda|^2}{\Lambda_\U^{2d_\U}}
\frac{m_t E_u^2}{2\pi} \left(
\frac{A_{d_\U}}{2\pi}\Delta^2(M_n^2)^{d_{\U}-2}\right),
  \end{split}
\end{equation}
with $M_n^2=m_t^2-2m_tE_u$, in agreement with Ref.~\cite{Georgi:2007ek}.
We see that each of the deconstructing particles $\lambda_n$
couples weaker and weaker as $\Delta\to0$ but their number in a fixed
interval of energies $dE_u$ is increasing inversely proportionally to
their coupling leading to finite $d\Gamma/dE_u$ in the scaling limit
$\Delta\to0$.

\section{Interference}
\label{sec:interference}

Another example considered in Ref.~\cite{Georgi:2007si} is the
coupling of a vector unparticle operator $\O^\mu$ to a neutral vector
or axial
lepton current, e.g., ($\ell=e$ or $\mu$)
\begin{equation}
  \label{eq:cvca}
c_{A\U}M_Z^{1-d_\U}\ \bar \ell \gamma^\mu\gamma_5 \ell\ \O_\mu,
\end{equation}
where, following notations in Ref.~\cite{Georgi:2007si} we expressed
the dimensionful coupling in units of the $Z$-boson mass $M_Z$.
These couplings produce contributions to, e.g., $e^+e^-\to\mu^+\mu^-$ 
amplitudes due to virtual unparticle which interferes with the Standard
Model $\gamma$ and $Z$ boson amplitudes. In the case of the vector
operator $\O^\mu$, which we assume to be conserved $\pd_\mu\O^\mu=0$,
the Eqs.~(\ref{eq:OO}), (\ref{eq:rhoO-scaling}), etc.\ 
generalize as
\begin{equation}
  \begin{split}
  \label{eq:OOmu}
    &  \Pi^{\mu\nu}(q)\equiv \int d^4x e^{iqx}\<0|T\O^\mu(x)\O^\nu(0)|0\>  \\
&= (-g^{\mu\nu} + q^\mu q^\nu/q^2)
\int \frac{dM^2}{2\pi} \rho_\O(M^2) \frac{i}{q^2-M^2+i\varepsilon}
  \end{split}
\end{equation}
where again by scale invariance $\rho_O$ is given by
Eq.~(\ref{eq:rhoO-scaling}). Evaluating the integral over $M^2$
one finds
\begin{equation}
  \label{eq:TOOmu}
\Pi^{\mu\nu}(q)
=
(-g^{\mu\nu} + q^\mu q^\nu/q^2)\frac{iA_{d_\U}}{2\sin(\pi d_\U)}
(-q^2-i\varepsilon)^{d_\U-2}.
\end{equation}
Deconstructing the unparticle operator $\O^\mu$ proceeds similarly to
the
scalar operator. We introduce decay constants $F_n$ via
\begin{equation}
  \label{eq:Fn-vec}
   \<0|\O^\mu(0)|\lambda_n\>=\epsilon^\mu F_n,
\end{equation}
where $\epsilon^\mu$ is the polarization of the massive vector
particle~$\lambda_n$. Then the correlation function is given by
\begin{equation}
  \label{eq:OO-Fn-vec}
\Pi^{\mu\nu}(q)
= (-g^{\mu\nu} + q^\mu q^\nu/q^2)\sum_n \frac{iF_n^2}{q^2-M_n^2+i\varepsilon}
\end{equation}
If we assume the same mass spectrum as in
Eq.~(\ref{eq:Mn}), the constants $F_n$ are again given by
Eq.~(\ref{eq:Fn-Delta}).

The contribution to the $e^+e^-\to\mu^+\mu^-$ amplitude
from the unparticle is proportional to the correlation function
(\ref{eq:OOmu}) and following Ref.~\cite{Georgi:2007si} we define:
\begin{equation}
  \label{eq:Delta_U}
  \Delta_\U = \frac{A_{d_\U}}{2\sin(\pi d_\U)}(-q^2-i\varepsilon)^{d_\U-2}.
\end{equation}
This amplitude interferes with the amplitude due to the virtual $Z$
proportional to
\begin{equation}
  \label{eq:DeltaZ}
  \Delta_Z = \frac1{q^2-M_Z^2+iM_Z\Gamma_Z}.
\end{equation}
This Standard Model amplitude is mostly real away from the $Z$-pole and is
mostly imaginary near the pole. The unusual property
of the unparticle amplitude (\ref{eq:Delta_U}) pointed out in
Ref.\cite{Georgi:2007si} is that it has nonzero imaginary part for all
$q^2>0$.  This allows the amplitudes $\Delta_\U$ and
$\Delta_Z$ to interfere even at the $Z$ pole, where the latter is imaginary.

This property follows naturally from the
deconstructed picture in which
\begin{equation}
  \label{eq:Delta_U-Fn-sum}
  \Delta_\U = \sum_n \frac{F_n^2}{q^2-M_n^2+i\varepsilon}.
\end{equation}

 The imaginary part of the 
amplitude $\Delta_\U$ as a function of
$q^2$ consists of a series of $\delta$-function peaks at $q^2=M_n^2$:
\begin{equation}
  \label{eq:peaks}
  \Im\Delta_\U = -\sum_n F_n^2 \pi\delta(q^2-M_n^2).
\end{equation}
Each peak becomes lower as
$F_n^2\sim\Delta^2\to0$, but their density increases.
 Converting the sum over $n$
into the integral over $M_n^2$ we find that
\begin{equation}
  \label{eq:Delta_U-Fn}
  \Im\Delta_\U \to - \frac {F_n^2}{\Delta^2} \,\pi 
= -\frac{A_{d_\U}}{2}(M_n^2)^{d_{\U}-2}
\end{equation}
in agreement with (\ref{eq:Delta_U}). The factor $\sin(\pi d_\U)$
which cancels in (\ref{eq:Delta_U}) never appears in the first place
in (\ref{eq:Delta_U-Fn}).

Away from the $Z$ pole, where $\Delta_Z$ is real, the interference
term is proportional to $\Re\Delta_\U$. This is given by the sum in 
(\ref{eq:Delta_U-Fn-sum}) where particles with masses $M_n^2<q^2$
contribute with the opposite sign from those  with $M_n^2>q^2$.
The case $d_\U=3/2$ is special, as pointed out in
Ref.\cite{Georgi:2007si}: $\Re\Delta_\U\sim \cot(\pi d_\U)$ vanishes.
This has a simple meaning -- at this value of $d_\U$ particles with
$M_n^2$ above
$q^2$ exactly cancel contribution of particles below $q^2$ (for any
$q^2$). 
This is most clear from the integral representation:
\begin{equation}
  \label{eq:Delta_dM^2}
  \Re\Delta_\U = \Xint-_0^\infty \!dM^2 \frac {(M^2)^{d_\U-2}}{q^2-M^2}
\end{equation}
That this (principal value) integral vanishes at $d_\U=3/2$ can be
seen by doing the change of variables $M\to q^2/M$ (mass inversion) 
which maps the regions above and below $q^2$ onto each other.

\section{Decay?}

We observe~(\ref{eq:Fn-Delta}) that each deconstructing
particle~$\lambda_n$ couples with strength proportional to
$F_n^2\sim\Delta^2$ which vanishes as $\Delta\to0$. Thus, in a certain
sense,  a true \mbox{($\Delta=0$)} unparticle, once produced, never
decays. 
This limiting procedure explains the apparent paradox
pointed out in Ref.~\cite{Georgi:2007si}: the finite imaginary part
(\ref{eq:Delta_U-Fn}) of the ``propagator'' of unparticle does not
mean it has finite lifetime.


What if the unparticle sector is almost conformal with a very small
but nonzero $\Delta$? The lifetime of a deconstructing particle
$\lambda_n$ is proportional to $F_n^{-2}\sim\Delta^{-2}$, and let us
assume that it is in the range that one can observe the displaced
vertex of $\lambda_n$ decay into ordinary Standard Model particles.
What would the signatures of such decays be?
For simplicity, let us assume here no interference with Standard Model
amplitudes.

First of all, the invariant mass spectrum of the decay products (e.g.,
lepton pairs) will not peak but will be a monotonous distribution (we
assume that $\Delta$ is much less than the experimental resolution).
Furthermore, the lifetime  would be proportional to
$F_n^{-2}$, which depends on~$M_n$ according to
Eq.~(\ref{eq:Fn-Delta}) or (\ref{eq:Fn-gamma}).  
There are of course trivial kinematic
and coupling factors, which might add an integer power of
$M_n$. One would therefore observe secondary vertices whose average
displacement is correlated with the invariant mass of the
products of decay.

For example, the contribution of the interaction (\ref{eq:cvca}) to
the decay rate of $\lambda_n$ is 
(taking $F_n$ from Eq.~(\ref{eq:Fn-Delta}))
\begin{equation}
  \begin{split}
  \label{eq:Gamma_n}
  \Gamma(\lambda_n\to \mu^+\mu^-) &= \frac{c_{A\U}^2 M_Z^{2-2d_\U} }{8\pi}\
  F_n^2\ M_n \\
&=
\frac{c_{A\U}^2 M_Z^{2-2d_\U} A_{d_\U}}{16\pi^2}\
  \Delta^2\ M_n^{2d_\U-3}.
  \end{split}
\end{equation}
Thus, the lifetime $\tau_{\rm d}$ measured through the mean displacement of a
vertex ($\tau_d=\ell_d/(\gamma v)$), if observed, will scale with the
reconstructed invariant mass as 
\begin{equation}
  \label{eq:displacement}
  \tau_{\rm d}=1/\Gamma\sim M^{3-2d_\U}.
\end{equation}

\section{Modeling and deconstructing unparticle using A\lowercase{d}S$_5$}
\label{sec:model-deconstr-unpar}

So far we have viewed deconstruction as an abstract
mathematical trick to cast the interaction of unparticle as a sum of
the interactions of the particles $\lambda_n$. This construction can
be made more explicit by considering a model of the unparticle based
on a 5-dimensional field theory. The idea is simple: restricting the
extent of the 5th dimension to size of order $1/\Delta$ 
will lead to the necessary discrete spectrum of deconstructing
particles~$\lambda_n$.

For concreteness and simplicity let us focus on a scalar unparticle
as in Section~\ref{sec:production}. The correlator (\ref{eq:OO-Fn})
can be obtained from a two-point Green's function of a massive
scalar field $\Phi(x,z)$, where $x$ is a Minkowski coordinate,
while the 5th coordinate $z$ can be thought of either as 
a continuous index, or as a 5th coordinate. We shall take AdS metric for this
5-dimensional space: $ds^2=(dx_\mu dx^\mu -dz^2)/z^2$. 
The Lagrangian (density in Minkowski space) reads:
\begin{equation}
  \label{eq:L5}
  {\cal L} =
 \int dz \sqrt{g}\ \left[g^{MN}\pd_M\Phi\pd_N\Phi - m_5^2\Phi^2\right]/2,
\end{equation}
where, as usual, $x^M=(x^1,x^2,x^3,t,z)$ and $g=\det||g_{MN}||$, $g_{tt}=+1$.
Note that the mass parameter $m_5$ is dimensionless, and so is the field
$\Phi$. The operator $\O$ can then be defined in terms of the field
$\Phi$ as follows:
\begin{equation}
  \label{eq:O-Phi}
  \O(x)\equiv \lim_{z\to0} z^{-d_\U}\Phi(x,z).
\end{equation}
In other words, the Standard-Model operators such as, e.g.,
$\pd_\mu(\bar u \gamma^\mu(1-\gamma_5)t)$ from Eq.~(\ref{eq:utU})
couple to the field $\Phi(x,z)$ only on the boundary $z\to0$. 

The dimension of the operator $d_\U$ determines
the required mass of the field (or is determined if the mass is given)
by the well-known formula:
\begin{equation}
  \label{eq:d_u-m_5}
  m_5^2 = d_\U(d_\U-4).
\end{equation}

The following analysis of this model  bears obvious
resemblance to the holographic technique described in
\cite{Gubser:1998bc,Witten:1998qj} and developed in many
subsequent works. In fact, our 5d model could be
perceived as a dual description of some 4-dimensional conformal field
theory in the sense of the AdS/CFT
correspondence~\cite{Maldacena:1997re}.
There is a similarity with the extra-dimensional scenarios
\cite{Arkani-Hamed:1998rs,Randall:1999ee}, but here we consider a {\em
  scalar\/} field in the bulk, rather than gravity. Gauge
fields have been also extensively studied in AdS$_5$ (see, e.g.,
\cite{Pomarol:2000hp,Goldberger:2002cz,Randall:2002qr} and refs.
therein), as well as massless scalar field
\cite{Csaki:1998qr,Polchinski:2001tt,Boschi-Filho:2002vd,deTeramond:2005su}
in a similar setup, but different contexts.  Here we shall focus on
the case of the {\em massive\/} scalar field which will allow a
nontrivial scaling dimension $d_\U$. With the understanding that many
elements of the following analysis can be found in the above literature,
 we shall, nevertheless, carry the following discussion in a
self-contained manner.

To understand and derive the relationship (\ref{eq:d_u-m_5}) between
the rescaling factor in Eq.~(\ref{eq:O-Phi}) and the mass parameter
given by (\ref{eq:d_u-m_5}), let us recall that the two-point
correlation function of the field $\Phi$ which appears in
\begin{equation}
  \label{eq:OO-PhiPhi}
  \< \O(x)\O(0) \> = \lim_{z,z'\to0} 
z^{-d_\U}(z')^{-d_\U}   \< \Phi(x,z) \Phi(0,z') \>
\end{equation}
is the Green's function of the linear differential operator
obtained by taking two variational
derivatives of the Lagrangian (\ref{eq:L5}) w.r.t. $\Phi$:
\begin{equation}
  \label{eq:L-operator}
  \left[\pd_z z^{-3} \pd_z + z^{-3}q^2 - z^{-5}m_5^2\right]
G(q;z,z') = \delta(z-z'),
\end{equation}
where $G(q;z,z')$ is the Fourier transform of  $\< \Phi(x,z)
\Phi(0,z') \>$ w.r.t. $x$. The behavior of the Green's function at
small $z$ is easy to find by noticing that the term $q^2$ is
negligible for $z\ll q^{-1}$, and that the equation
(\ref{eq:L-operator}) with $q^2$ neglected is solved (for $z\neq z'$)
by a power ansatz $G\sim z^\sigma$, with
$\sigma(\sigma-4)=m_5^2$. Thus, on the account of
Eq.~(\ref{eq:d_u-m_5}): $G(q;z,z')\sim z^{d_\U} (z')^{d_\U}$.%
\footnote{Explicit solution, which we do not need here, is
$G(q;z,z')=
(\pi/2)(zz')^2J_{d_\U-2}(qz)Y_{d_\U-2}(qz)\theta(z'-z)+(z\leftrightarrow z')$.}
Thus the correlator
 $ \< \O(x)\O(0) \>$ in Eq.~(\ref{eq:OO-PhiPhi}) is finite in the
limit $z,z'\to0$, and by dimension counting must be proportional to
$x^{-2d_\U}$, which is what the dimension of the operator $\O(x)$ implies.

Consider now AdS space with finite extent in the 
$z$ direction (AdS slice):
$z\in[0,z_m]$, with $z_m\sim 1/\Delta$.
 We can write the representation for the Green's function
in terms of the normalizable modes:
\begin{equation}
  \label{eq:G-psipsi}
  G(q;z,z') = \sum_n\frac{\phi_n(z)\phi_n(z')}{q^2-M_n^2+i\varepsilon}.
\end{equation}
The normal modes $\phi_n$ are normalized according to 
\begin{equation}
  \label{eq:psi-norm}
  \int dz z^{-3}\phi_n^2=1.
\end{equation}
They behave as $\phi_n\sim z^{d_\U}$ for small $z$,
and the constants $F_n$ can be identified, comparing (\ref{eq:OO-Fn}),
(\ref{eq:OO-PhiPhi}) and (\ref{eq:G-psipsi}), as 
\begin{equation}
  \label{eq:Fn-psin}
  F_n = \lim_{z\to0} z^{-d_\U}\phi_n(z).
\end{equation}
This result is similar to the expression for the meson decay
constants in AdS/QCD
in terms of the $z\to0$ asymptotics of the normalizable 
modes~\cite{Erlich:2005qh,DaRold:2005zs}.

The distribution of masses $M_n$ can be obtained given the specific
boundary conditions at $z_m$. Instead of relying on the exact solution,
let us note that only $n\gg1$ modes interest us here, since, by assumption,
$\Delta\ll M_n$.  The equation for normal modes (\ref{eq:L-operator})
can be cast into Schr\"odinger form by substitution:
$\phi_n=z^{-3/2}\psi_n$, and then solved in the WKB approximation. For
$z\gg M_n^{-1}$, including $z=z_m$ boundary, the large $n$
modes are $\psi_n\sim \sin(M_nz-C_1)$, where constant $C_1$ depends on
$m_5$ (i.e., on $d_\U$), but not on $n$. Thus the mass spectrum is given by
$M_n \sim (z_m)^{-1}(\pi n+C_2)$, where constant $C_2$ is related to $C_1$
and depends also on the type of the boundary condition at $z_m$
(e.g., $C_2=C_1$ for Dirichlet boundary condition).
Neglecting ${\cal O}(1/n)$ terms we find quite generally
$M_n^2\to\Delta^2 n^2$ for $n\gg1$, where $\Delta=\pi/z_m$. 
To obtain the linear spectrum
as in~(\ref{eq:Mn}) one can modify the AdS background at large $z$,
 as it is done
in Ref.~\cite{Karch:2006pv},
instead of cutting the space off at~$z_m$.

The arguments in this section assume that the dimension
of the operator satisfies $1<d_\U<2$. Indeed, for the normal modes to
be normalizable in the sense of Eq.~(\ref{eq:psi-norm}) we must have
$d_\U>1$. More subtly, in order for the Green's function in
Eq.~(\ref{eq:L-operator}) to have the behavior $z^{d_\U} (z')^{d_U}$
as $z,z'\to 0$, the value $d_\U$ must be the {\em smallest\/} of the
two solutions of quadratic equation (\ref{eq:d_u-m_5}), which means
$d_\U<2$. The integer values $d_\U=1,2$ are of course special and, although
interesting, will not be considered here.  Discussion of the possible
dimensions of a scalar field in a conformal theory and in AdS/CFT
correspondence can be found, e.g., in
\cite{Balasubramanian:1998sn,Klebanov:1999tb}.

In summary, we have seen how the notion of unparticle can be somewhat
demystified by representing it as an infinite tower of massive
particles with controllable mass-squared spacing $\Delta^2$. We used
such a deconstruction technique to rederive and clarify the peculiar
properties of unparticle pointed out in
Refs.~\cite{Georgi:2007ek,Georgi:2007si} and to show that pure
($\Delta=0$) unparticle cannot decay, while for small but nonzero
$\Delta$ the decay is possible, with peculiar signature.
Finally, we described a possible field theory realization of the
deconstruction procedure using a slice of AdS$_5$ space. 

The author thanks Prof. W.~Y.~Keung for bringing
Ref.~\cite{Georgi:2007ek} to his attention and for
stimulating discussions. This research is supported by the
DOE grant No.\ DE-FG0201ER41195.

\end{document}